\documentstyle[12pt]{article}
\newcommand{\be}{\begin{equation}}
\newcommand{\bea}{\begin{eqnarray}}
\newcommand{\eea}{\end{eqnarray}}
\newcommand{\ba}{\begin{array}}
\newcommand{\ea}{\end{array}}
\newcommand{\ee}{\end{equation}}

\expandafter\ifx\csname mathbbm\endcsname\relax

\else

\fi
\begin{document}
\begin{titlepage}
\hfill
\vbox{
    \halign{#\hfil         \cr
           IPM/P-2002/050 \cr
           hep-th/0210122  \cr
           } % end of \halign
      }  % end of \vbox
\vspace*{20mm}
\begin{center}
{\Large {\bf Circular Semiclassical String solutions on Confining
AdS/CFT Backgrounds }\\ }

\vspace*{15mm}
\vspace*{1mm}
{Mohsen Alishahiha$^a$ \footnote{Alishah@theory.ipm.ac.ir} 
and Amir E. Mosaffa$^{a,b}$}
\footnote{Mosaffa@theory.ipm.ac.ir} \\
\vspace*{1cm}

{\it$^a$ Institute for Studies in Theoretical Physics 
and Mathematics (IPM)\\
P.O. Box 19395-5531, Tehran, Iran \\ \vspace{3mm}
$^b$ Department of Physics, Sharif University of Technology\\
P.O. Box 11365-9161, Tehran, Iran}\\

\vspace*{1cm}
%%\maketitle
\end{center}

\begin{abstract}
We study multiwrapped circular string pulsating in the radial 
direction of AdS black hole. We compute the energy of this string as
a function of a large quantum number $n$. One then could associate 
it with energy and a quantum number of states in the dual
finite temperature ${\cal N}=4$ SYM theory as well as three 
dimensional pure gauge theory. We observe that the $n$ dependence
of the energy has a universal form. We have also considered pulsating 
string in the background of the near-extremal D4-brane solution.
Circular pulsating membrane in M-theory on $AdS_7\times S^4$ 
has also been studied. 
\end{abstract}

\end{titlepage}

\section{Introduction}
It is conjectured that type IIB
string theory on $AdS_5\times S^5$ with $N$ fluxes on the 
$S^5$ is dual to four dimensional ${\cal N}=4$ $SU(N)$ SYM 
theory \cite{{MAL},{GKP1},{WI}}. According to this conjecture 
the spectrum of single string states on $AdS_5\times S^5$ 
corresponds to the spectrum of single trace operators of the 
${\cal N}=4$ gauge theory. However, until recently, this 
correspondence had only been studied for supergravity modes 
on $AdS_5\times S^5$ which are in one-to-one correspondence 
with the chiral operators of the ${\cal N}=4$ gauge theory. 

In an interesting recent development the authors of \cite{BMN}
have been able to study the AdS/CFT correspondence beyond the
gravity modes. In fact it has been conjectured \cite{BMN} that 
the string theory on 
the maximally supersymmetric ten-dimensional PP-wave has a 
description in terms of a certain subsector of the
large $N$ four-dimensional ${\cal N}=4\; SU(N)$ supersymmetric 
gauge theory at weak coupling. More precisely this subsector is 
parametrized by states with conformal weight $\Delta$ carrying 
$J$ units of charge under the $U(1)$ subgroup of the $SU(4)_{R}$ 
R-symmetry of the gauge theory, such that both $\Delta$ and $J$ 
are parametrically large in the large `t Hooft coupling while 
their difference, $\Delta-J$ is finite. Therefore it has been possible 
to work out the perturbative string spectrum from the gauge theory 
side. 

Shortly after, the authors of \cite{GKP} identified certain classical 
solutions representing highly 
excited string states carrying large angular momentum in the 
$AdS_5$ part of the metric with gauge theory operators with high 
spin $S$ and conformal dimension $\Delta$ which is identified  
with the classical energy of the solution in the global
$AdS$ coordinates. An interesting observation of \cite{GKP} is 
that the classical
energy of the rotating string in $AdS_5$ space in the limit of 
$S\gg \sqrt{\lambda}$ scales as 
\be
\Delta-S={\sqrt{\lambda}\over \pi}\ln{S\over \sqrt{\lambda}}
+\cdots \;,
\ee
which looks like the logarithmic growth of anomalous 
dimensions of operators with spin in the gauge theory. It has 
also been shown that the BMN  
operators \cite{BMN} can also be identified with classical
solutions of string in $AdS_5$ with angular momentum in $S^5$ 
space \cite{GKP}.

A generalization for the case where the string is stretched along
radial direction of $AdS$ and rotates 
along both $AdS_5$ and $S^5$ spaces has also been studied in 
\cite{FT}. This solution corresponds to those operators 
which have both spin and R-charge in the gauge theory side. 
A more general solution where the string is also
stretched in an angular coordinate of $S^5$ has been studied in 
\cite{Russo}. For further study in this direction see 
\cite{SEZ}-\cite{Flo}.

Recently another semiclassical string solution on $AdS_5\times S^5$
has been studied \cite{Min}. 
This string configuration corresponds to a 
multiwrapped circular string pulsating in the radial direction of $AdS_5$
and fixed to a point on $S^5$. More precisely in the notation in which
the AdS metric is written as 
\bea
ds^2&=&\alpha' \lambda^{1/2}(-\cosh^2\rho\;dt^2+d\rho^2+\sinh^2\rho\;
d\Omega^2_3)\;,\cr
d\Omega_3^2&=&\cos^2\theta\;d\psi^2+d\theta^2+\sin^2\theta\;d\phi^2\;,
\eea
the string solution is given by
\be
t=\tau,\;\;\;\;\;\rho=\rho(\tau),\;\;\;\;\;\phi=m\sigma,\;\;\;\;\;
\theta={\pi\over 2}\;,
\ee
where $\sigma$ and $\tau$ are worldsheet coordinates.

For the string with large values of energy, one can use the 
Bohr-Sommerfeld analysis to find the energy of the string as 
a function of a large quantum number, the excitation level 
of the string, as following\footnote{The leading term in
this expression was first obtained in \cite{DEL}.}
\be 
E\approx 2n+{8\pi^{1/2}\over \Gamma({1\over 2})^2}
\lambda^{1/4}\sqrt{mn}\;.
\label{MINR}
\ee

Using the AdS/CFT correspondence one can relate the quantum number
to the dimension of an
operator in the ${\cal N}=4$ SYM theory. From (\ref{MINR})
one can read the anomalous dimension of the
corresponding gauge theory operator \cite{Min}
\be
\Delta\sim\Delta_0+\lambda^{1/4}\sqrt{mn}\;,
\ee
where $\Delta_0=2n$ is the bare dimension of the operator and
$mn$ corresponds to the string level. 

One can also consider a circular string pulsating in the $S^5$.
In this case the energy as a function of a large quantum
number is given by \cite{Min}
\be
\Delta\sim\Delta_0+\lambda \frac{m^2}{2n}\;.
\ee

The aim of this article is to generalize this string solution 
to the AdS black hole background. Since this background is 
conjectured to be dual to a confining gauge theory, this might increase 
our knowledge about this theory and thereby QCD.

The paper is organized as follows. In section 2 we shall consider
circular pulsating string in $AdS_5$ black hole. Using WKB approximation
we will compute the energy of string in terms of a large quantum
number. This string could be identified with some state in 
finite temperature ${\cal N}=4$ SYM theory.
We also observe that the energy dependence of $n$ has a universal
form for ${\cal N}=4$ SYM theory and its deformations. In section 3
we will study  circular pulsating string in the background of near-extremal 
D4-brane solution in the decoupling limit. This is supposed to be dual to 
a four dimensional pure gauge theory (QCD)\cite{Witten} \footnote{We note, 
however that at low energies this theory has more state with the same mass 
than pure gauge theory in four dimensions \cite{ORT}.}. In section 4 we
shall study circular pulsating membrane solution in M-theory on 
$AdS_7\times S^4$. The last section is devoted to conclusions.

\section{Circular string in confining gravity background} 

In this section following \cite{Min} we shall study the circular
string in the background which describes confining theories. 
In the AdS/CFT context these backgrounds have the following typical 
form  
\be
ds^2=f(r)\;dM_d^2+dr^2+dN_{9-d}^2\;,
\ee
where $dM_d^2$ is the Minkowski metric of $R^d$ where the gauge theory 
lives, $r$ is holographic radial coordinate and $dN_{9-d}^2$ is 
the internal space of full 10-dimensional superstring theory.

The gauge theory has confinement phase if $f(r)$ has a local minimum at,
say, $r=r_*$ with $f(r_*)>0$ (for detail see for example \cite{Sonn}). 
In this case the Wilson loop computation leads
to a linear potential for quark-anitquark and the effective string tension
is given by
\be
T_0={f(r_*)\over 2\pi \alpha'}\;.
\ee    

In this section we study those deformed $AdS_5$ gravity backgrounds which
exhibit this property. 

\subsection{Finite temperature}
Let us first consider the finite temperature ${\cal N}=4$ SYM 
theory. The gravity dual of this theory is given by AdS black hole 
with the metric \cite{{de},{KM},{Witten}}
\bea
ds^2&=&-f(r)\;dt^2+f^{-1}(r)\;dr^2+r^2d\Omega^2_3+R^2d\Omega_5^2,\cr
&&\cr 
d\Omega_3^2&=&d\theta^2+\sin^2\theta\; d\phi^2+\cos^2\theta\;d\psi^2\;,
\label{FIN}
\eea
with
\be
f(r)=1+{r^2\over R^2}-{M^2\over r^2}\;,
\ee
where $M$ is related to the black hole mass and for high-temperature
we have $M^2/R^2\sim (RT)^4$ with $T$ being the temperature \cite{Witten}.
This metric is asymptotic to $AdS_5$ in global coordinates and
the boundary of this gravity solution where the gauge theory lives is 
$S^1\times S^3$. This solution provides gravity description of the finite
temperature of ${\cal N}=4$ SYM theory on $S^1\times S^3$. 

Semiclassical description of rotating string in this background has been
studied in \cite{BAR}. Now we would like to study
the semiclassical quantization of a circular string that expands and 
contracts in this background. In other words we shall consider a 
pulsating string which is wrapped $m$ times around the $\phi$ direction
sitting at $\theta={\pi\over 2}$. The string configuration is given by
\be 
t=\tau,\;\;\;\;\;\; r=r(\tau),\;\;\;\;\;\; \phi=m \sigma\;\;\;\;\;
\theta={\pi\over 2}\;,
\label{SS}
\ee 
where the string worldsheet is parameterized by $\tau$ and $\sigma$.

The Nambu-Goto action
\be
S=-{1\over 2\pi \alpha'}\int d\tau d\sigma\; \sqrt{-\det(G_{\mu\nu}
\partial_{\alpha}X^{\mu}\partial_{\beta}X^{\nu})}
\ee
for the configuration (\ref{SS}) reads
\be
S=-{m\over \alpha'}\int dt\; (r^2/f)^{1/2}\sqrt{f^2-{\dot r}^2}\;.
\label{act}
\ee
Here dot represents derivative with respect to $t$. By making use of 
the following change of variable 
\be
\xi=\int {dr\over f(r)}\;,
\ee 
the action (\ref{act}) can be recast to
\be
S=-{m\over \alpha'}\int dt\; g(\xi)\sqrt{1-{\dot \xi}^2}\;,
\ee
where $g(\xi)=\sqrt{r^2f}$. Treating this action as a one-dimensional
quantum mechanical system, one finds the Hamiltonian of the theory as 
following
\be
H=\sqrt{\Pi^2+\left({m\over \alpha'}\right)^2g(\xi)^2}\;, 
\ee
with
\be
\Pi={m\over \alpha'}\;g(\xi)^2{{\dot \xi}\over \sqrt{1-{\dot \xi}^2}}
\ee
being the canonical momentum. Note that $H^2$ can be considered as a one
dimensional quantum mechanical system with potential 
\be
V(\xi)=\left({m\over \alpha'}\right)g(\xi)^2\;.
\ee

Therefore, we can use the Bohr-Sommerfeld analysis for the quantization
of the states.  The quantization condition is given by
\be
(n+{1\over 2})\pi=\int_{\xi_1}^{\xi_2}d\xi\;\sqrt{E^2-
\left({m\over \alpha'}\right)^2g(\xi)^2}\;, 
\ee
where $\xi_{1,2}$ are the turning points. 

It is useful to return to the original coordinate $r$ in which the
quantization condition becomes
\be
(n+{1\over 2})\pi=E\left[\int_{r_H}^{r_1}{dr\over 1+{r^2\over R^2}
-{M^2\over r^2}}-\int_{r_H}^{r_1}dr {1-\sqrt{1-{1\over B^2}\left(r^2+{r^4
\over R^2}
-M^2\right)}\over 1+{r^2\over R^2}
-{M^2\over r^2}}\right]\;,
\label{INTT}
\ee
where $B=\alpha' E/m$ and 
\bea
r_{1}&=&{R\over \sqrt{2}}\left(\sqrt{1+{4(B^2+M^2)\over R^2}}
-1\right)^{1/2}\;,\cr &&\cr
r_H&=&{R\over \sqrt{2}}\left(\sqrt{1+{4M^2\over R^2}}
-1\right)^{1/2}\;.
\eea

For large $B$ ($B/R \gg 1$) the first integral in (\ref{INTT}) 
becomes
\be 
\int_{r_H}^{r_1}{dr\over 1+{r^2\over R^2}
-{M^2\over r^2}}\approx {\pi\over 2}\sqrt{{R^3\over M}}-
\sqrt{{R^3\over B}}\;,
\ee
while for second integral, setting $r=y\sqrt{BR}$, one finds
\bea
\int_{r_H}^{r_1}dr {1-\sqrt{1-{1\over B^2}\left(r^2+{r^4\over R^2}
-M^2\right)}\over 1+{r^2\over R^2}
-{M^2\over r^2}}&\stackrel{{\rm large}\;B}{\longrightarrow}&
\sqrt{{R^3\over B}}\int_0^1{dy\over y^2}(1-\sqrt{1-y^4})\cr
&&\cr
&=&\sqrt{{R^3\over B}}\left(-1+{(2\pi)^{3/2}\over\Gamma({1\over 4})^2}
\right)
\eea
Thus altogether we get
\be
(n+{1\over 2})\pi\approx R\bigg{[}{\pi\over 2}\left({R\over M}\right)^{1/2}
E-{(2\pi)^{3/2}\over\Gamma({1\over 4})^2} \left({MR\over \alpha'}
\right)^{1/2}\sqrt{E}\bigg{]}\;,
\ee
which can be inverted to find energy as a function of $n$
\be
E R\approx \left({4M\over R}\right)^{1/2}n+{4\sqrt{\pi}\over 
\Gamma({1\over 4})^2}
\sqrt{{4MR\over \alpha'}}\;\left({4M\over R}\right)^{1/4}
\sqrt{mn}\;.
\ee
As we see the $n$ dependence of energy is the same as that
in the conformal case (\ref{MINR}), of course, up to a numerical
coefficient. We note, however, its interpretation from finite 
temperature ${\cal N}=4$ SYM theory point of view might be 
different. This is because the notion of anomalous dimension is 
defined only at or near a conformal point. 
Nevertheless this expression can be thought of as the
dispersion relation of stationary states in the gauge theory
side.

\subsection{Witten's confining model}

Starting from the finite temperature solution (\ref{FIN}) for large 
$M$ one can use a change of variable which
reduces the solution (\ref{FIN}) to a solution with
boundary $R^3\times S^1$. This solution would provide gravity 
description of three dimensional gauge theory which exhibits 
confinement. The resulting solution is the same as that obtained
by scaling the near-extremal brane solution 
\cite{{HR},{IMSY}}.

For the model we are considering the corresponding solution is
obtained from the near-extremal D3-brane solution which is given by
\be
ds^2=r^2\left(R^2\;h(r)\;d\phi^2\;+\;dy^2\;-\;
dt^2\;\right)+{dr^2\over r^2h(r)}\;+\;d\Omega_5^2\;, 
\ee 
where $h(r)=1-({r_0\over r})^4$.

The rotating string in this background has recently been studied 
in \cite{BAR2}. We would like to study the pulsating string in this 
background which is also wrapped around the $\phi$ direction. 
This string configuration is given by
\be 
t=\tau,\;\;\;\;\;r=r(\tau),\;\;\;\;\;\phi=m\sigma\;.
\ee 
The Nambu-Goto action (\ref{act}) for this 
solution reads 
\be 
S={-mR\over\alpha'}\int\;dt\;\sqrt{h\; r^4-{\dot r}^2}\;.
\ee 
Defining a new variable, $\xi$, such that 
${dr\over d\xi}=\sqrt{r^4-r_0^4}$, and  using the same procedure as in 
the previous subsection, we can write the Hamiltonian in the following 
form
\be
H=\sqrt{\Pi^2+({m R\over \alpha'})^2\;({dr\over d\xi})^2}\;. 
\ee 
One can now use the Bohr-Sommerfeld analysis for the quantization
of the states. The quantization condition in the original $r$
coordinate is given by
\be
(n+{1\over 2})\;\pi=E\bigg{[}\;\int_{r_0}^{r_1}\;{dr\over 
\sqrt{r^4-r_0^4}}\;-\int_{r_0}^{r_1} dr {1-\sqrt{
{1-{1\over B^4}(r^4-r_0^4)}}\over \sqrt{r^4-r_0^4}}\;\;\bigg{]}\;, 
\label{BSC3}
\ee 
where $B^2=\alpha' E/mR$ and $r_1=B(1+r_0^4/B^4)^{1/4}$.

For large $B$ ($B/r_0 \gg 1$) the first integral (\ref{BSC3})
becomes
\be 
\int_{r_0}^{r_1}\;{dr\over 
\sqrt{r^4-r_0^4}}={1\over4}\;\pi^{1/2}\;{\Gamma({1\over 4})\over 
\Gamma({3\over4})}\;{1\over r_0}-\;{1\over B}+O({1\over B^2})\;,
\ee 
while for the second one, setting $y=B^{-1}\;r$, we find  
\be 
{1\over B}\int_{{r_0\over B}}^1\; {dy\over\sqrt{y^4-B^{-4}r_0^4}} 
\left(1-\sqrt{1-y^4+r_0^4/B^4}\right)\;, 
\ee 
which leads to
\be 
{1\over B}\int_0^1{dy\over y^2}\;(1-\sqrt{1-y^4})={1\over 
B}\left(-1+{(2\pi)^{3/2}\over\Gamma({1\over 4})^2}\right)\;.
\ee 
Therefore we find the following quantization condition at large 
$B$ limit 
\be 
(n+{1\over 2})\;\pi\approx {1\over4}\;\pi^{1/2}\;{\Gamma({1\over 4})\over 
\Gamma({3\over4})}\;{E\over r_0}-{(2\pi)^{3/2}\over 
\Gamma({1\over4})^2}\;({mR\over \alpha'})^{1/2}\;\sqrt{E}\;, 
\ee 
which can be inverted to find the energy as a function of $n$ 
\be 
E\approx 4\;\pi^{1/2}\;{\Gamma({3\over4})\over \Gamma({1\over 4})}\;r_0
\;n\;\left[\;1+\;8^{3/2}\pi^{3/4}\;{\Gamma({3\over 
4})^{3/2}\over\Gamma({1\over 4})^{7/2}}\;\left({mRr_0\over \alpha' 
\;n}\right)^{1/2}\;\right]\;.
\ee
Note that the energy as a function of $n$ has the same form as
that in the previous cases for both conformal and finite 
temperature models where $\sqrt{M/R}$ plays the role of $r_0$.
Therefore one might conclude that the $n$ dependence of energy 
has a universal form. Of course, as we
will see, it does depend on the dimension of the theory.

\section{Circular string in gravity dual of QCD}

In the context of AdS/CFT correspondence the gravity dual of 
QCD has been proposed in \cite{Witten} and further studied
in \cite{OO}. The corresponding gravity solution is obtained 
from near-extremal D4-brane solution \cite{IMSY} which can be
recast to 
\bea
ds^2&=&r^{3/2}\left(R^2\;h(r)\;d\phi^2\;+\;d{\vec y}_3^2\;-\;
dt^2\;\right)+{dr^2\over r^{3/2}h(r)}\;+r^{1/2}d\Omega_4^2\;, \cr
h(r)&=&1-({r_0\over r})^3\;.
\label{NED4}
\eea
This background has been used to study several properties of 
pure four-dimensional Yang-Mills theory including 
quark-anitquark potential, glueballs masses etc \cite{OO}-
\cite{HO}. These results are qualitatively in agreement
with what is expected from QCD, though it is, by now, known that
in low energies this background gives KK spectrums with the 
same masses as the glueballs masses and therefore has more states 
than there are in a pure gauge theory \cite{ORT}. Nevertheless 
using 
this background we would expect to get, at least, some 
qualitative results by studying semiclassical string in this
background. The rotating string in this background has recently 
been studied in \cite{BAR2}. We shall study a pulsating string in 
this  background which is also wrapped around the $\phi$ direction. 
This string configuration is given by
\be 
t=\tau,\;\;\;\;\;r=r(\tau),\;\;\;\;\;\phi=m\sigma\;.
\ee 
The Nambu-Goto action (\ref{act}) for this 
solution reads 
\be 
S={-mR\over\alpha'}\int\;dt\;\sqrt{r^3h-{\dot r}^2}\;. 
\ee 
By making use of a change of variable such that 
${dr \over d\xi}=\sqrt{r^3-r_0^3}$, the Hamiltonian can be written
as following 
\be
H=\sqrt{\Pi^2+({m R\over \alpha'})^2\;({dr\over d\xi})^2}\;, 
\ee 
which leads to the following quantization condition in the 
original $r$ coordinate due to the Bohr-Sommerfeld analysis
\be
(n+{1\over 2})\;\pi=E\bigg{[}\;\int_{r_0}^{r_1}\;{dr\over 
\sqrt{r^3-r_0^3}}\;-\int_{r_0}^{r_1} dr {1-\sqrt{
{1-{1\over B^3}(r^3-r_0^3)}}\over \sqrt{r^3-r_0^3}}\;\;\bigg{]}\;, 
\label{BSC4}
\ee 
where $B^{3/2}=\alpha' E/mR$ and $r_1=B(1+r_0^3/B^3)^{1/3}$.

One can perform the integrals in (\ref{BSC4}) to find the 
dependence of energy on $n$. Doing so we get
\be
E\approx 3\sqrt{\pi}{\Gamma({2\over 3})\over \Gamma({1\over 6})}
\sqrt{r_0n^2}\bigg{[}1+{1\over \pi^{1/6}}\left({3\Gamma({2\over 3})
\over \Gamma({1\over 6})}\right)^{2/3}\left({mRr_0\over n\alpha'}
\right)^{1/3}\bigg{]}\;.
\ee
We note that although the $n$ dependence of the energy is not the
same as in the previous case, it belongs to another universality class.
This can be understood from the fact that the corresponding
gravity background (\ref{NED4}) can be obtained from 
the $AdS_7$ black hole compactified on a circle. Therefore
we would expect to find a universal form for the energy as 
a function of $n$ with the same form as that which could 
have been obtained from circular pulsating membrane in the
M-theory on $AdS_7\times S^4$. To see this, in next section we 
shall study this membrane configuration in M-theory.

\section{Circular membrane in M-theory $AdS_7\times S^4$ background}

In this section we study multiwrapped circular membrane
pulsating in the radial direction of $AdS_7$ in M-theory.
Let us start with the gravity solution of 
$AdS_7\times S^4$ in the global coordinates 
\bea 
l_p^{-2}dS^2&=& 
4R^2\bigg{[}-dt^2\cosh^2\rho+d\rho^2+ \sinh^2\rho\left(d\psi_1^2
+\cos^2\psi_1\; d\psi_2^2+\sin^2\psi_1\;d\Omega_3^2\right)\cr 
&+&{1\over 4}\left(d\alpha^2+\cos^2\alpha\;d\theta^2+\sin^2\alpha\;
(d\beta^2+\cos^2\beta\; d\gamma^2)\right)\bigg{]}\;, 
\cr d\Omega^2_3 &=&d\psi_3^2+\cos^2\psi_3\; 
d\psi_4^2+\cos^2\psi_3\;\cos^2\psi_4\;d\psi_5^2,
\label{ADSBAC}
\eea
where 
$R^3=\pi N$. Note that,  we are using a unit in which $R$ is 
dimensionless. The supersymmetric action of the M-theory
supermembrane has been studied in \cite{BST}.
Here we shall only consider the 
bosonic part of the action which can be written as 
following
\be
I=-{1\over (2\pi)^2l_p^3}\int 
d\xi^3\left(\sqrt{-\det(G_{\mu\nu}\partial_{a}x^{\mu}\partial_b 
x^{\nu})}+{1\over 
6}\epsilon^{ijk}\partial_ix^{\mu}\partial_jx^{\nu}\partial_k 
x^{\lambda}C_{\mu\nu\lambda} \right), 
\label{M2ACT}
 \ee 
where$(\xi_1,\xi_2,\xi_3)=(\tau,\delta,\sigma)$ are coordinates 
which parameterize the membrane worldvolume. $x^{\mu},\;\mu=0,\cdots,10$ 
are space-time coordinates and $C_{\mu\nu\lambda}$ is the massless 
M-theory three-form. 
The rotating membrane solution in this background has been
studied in \cite{{SEZ},{AliGh}}. We now look for a soliton 
solution representing a multiwrapped membrane pulsating in 
the radial coordinate which is fixed to a point on $S^4$. 
The solution is given by
\be
t=\tau,\;\;\;\;\psi_2=\sqrt{2}a\delta,\;\;\;\; 
\psi_5=\sqrt{2}m \sigma,\;\;\;\; \rho=\rho(\tau),\;\;\;\; 
\psi_1={\pi\over 4},
\label{MEMSOL}
\ee
all other coordinates are set to 
zero. For the solution (\ref{MEMSOL}) the CS part of the 
membrane action (\ref{M2ACT}) is zero and therefore the membrane 
action (\ref{M2ACT}) reads 
\be 
I=-(2R)^3 a m \int dt\; 
\sinh^2\rho\;\sqrt{\cosh^2\rho-{\dot \rho}^2}\;.
\label{ACTION}
\ee
Introducing a new variable $\xi=\arcsin(\tanh\rho)$,
the action (\ref{ACTION}) gets the following form 
\be 
I=-(2R)^3 a m\int dt\; \tan^2\xi\;\sec\xi\;
\sqrt{1-\dot \xi^2}\;. 
\ee 
In these coordinates the Hamiltonian is
\be 
H=\Pi\;\dot\xi-L=\sqrt{\Pi^2+(2R)^6 a^2m^2 
\tan^4\xi \sec^2\xi}\;,
\ee 
which can be treated as a one dimensional quantum 
mechanical system such that the potential for $H^2$ is
given by  
\be
V(\xi)=2 (2R)^6 a^2 m^2 \tan^4\xi 
\sec^2\xi 
\ee 
To find the energy levels of the string states one can make use 
of the Bohr-Sommerfeld analysis. Since $\xi$ has to be positive we 
can symmetrize the potential and only consider the even states. 
Thus one gets 
\be 
(2n+{1\over 2})\pi=\int_{-\xi_0}^{\xi_0}
d\xi\;\sqrt{E^2-(2R)^6 a^2m^2 \tan^4\xi 
\sec^2\xi}\;,
\label{eeee}
\ee
where $\pm\xi_0$ are the turning points which solve the following 
equation
\be
E^2=(2R)^6\;a^2\;m^2\;\tan^4{\xi_0}\;\sec^2{\xi_0}\;.
\ee

Defining $B={E\over(2R)^6a^2 m^2}$ and setting 
$ \tan\xi=\;B^{1/3}\; y$, the equation (\ref{eeee}) can be 
recast to
\be 
(2n+{1\over 2})\pi=2 (2R)^3 a mB^{2/3} 
\bigg{[}\int_0^{y_0} { dy \over (B^{-2/3}+y^2)} - \int_0^{y_0} dy\; 
{1-\sqrt{1-y^4/B^{2/3}-y^6}\over B^{-2/3}+y^2 }\bigg{]}\;,
\ee
where $y_0$ is the 
turning point in the new coordinates. In the limit  
$B\rightarrow\infty$ and for large energies, we get
\be
E\approx\;2n\;+{1.12\over\pi}\;2^{8/3}\;R\;(n\;
\sqrt{a m})^{2/3}\;,
\label{CIM2}
\ee
which has the same $n$ dependence as that we found in the 
previous section for the near-extremal D4-brane. This also
shows that the $n$ dependence of the energy has a universal
form.

Note also that since the gravity solution (\ref{ADSBAC}) is conjectured
to be dual to the (0,2) theory one can use the result (\ref{CIM2}) to
study the spectrum of this theory. In fcat, from the $(0,2)$ theory
point of view this means that there exist operators with the following 
anomalous dimension 
\be
\Delta\sim \Delta_0+ R (n \sqrt{am})^{2/3}\;.
\label{M02}
\ee

\section{Conclusion}

In this paper, we have studied a new class of semiclassical string 
solution in various backgrounds in string theory which 
are conjectured to be dual to the non-supersymmetric confining
gauge theories. This solution represents a multiwrapped circular
string pulsating in the radial coordinate of the gravity
solution of the near-extremal D3 and D4-branes as well as
$AdS_5$ black hole.

We have evaluated the energy of this string configuration
as a function of a large quantum number. We observe that the
result is universal, i.e. is the same for conformal,
non-conformal and non-supersymmetric cases. Of course the result 
does depend on the dimension of the theory. For example for
${\cal N}=4$ and its deformations, we find a universal form as
following
\be
E\sim n+f(\lambda) \sqrt{mn}\;,
\ee
where $\lambda$ is `t Hooft effective coupling.
We note, however, that their interpretations from different gauge 
theories points 
of view might be different. This is because the notion 
of anomalous dimension is defined only at or near a conformal
point. Nevertheless this expression can be thought of  as the
dispersion relation of stationary states in the gauge theory
side.

One could also consider those string solutions which pulsate on 
the internal part of the full string background, {\it e.g.}
$S^5$ or $S^4$. But since the $n$ dependence of energy
would have a universal form, we would expect to find the same
result as that in the conformal case. For example for deformed
$AdS_5$ background it is given by 
\be
E\sim 2n+\lambda \frac{m^2}{2n}\;.
\ee

We have also studied a multiwrapped circular
membrane pulsating in the radial coordinate of $AdS_7$ in 
M-theory. We have then been able to find the energy dependence
on a large quantum number $n$. We note that the $n$ dependence
of energy is in the the same universality class as that for 
the pulsating string in the background of
the near-extremal D4-brane. This can be understood from the fact
that the near-extremal D4-brane solution can be obtained from $AdS_7$ 
black hole compactified on a circle. 

On the other hand since the gravity solution (\ref{ADSBAC}) is 
dual to the (0,2) theory one can use the result (\ref{CIM2}) to
study the spectrum of this theory. In fact, this result is our
prediction of existence of an operator in (0,2) theory with 
anomalous dimension given by (\ref{M02}). It would be interesting
to find the explicit form of the operator in the (0,2) theory.

\vspace*{.4cm}

{ \bf Acknowledgements}

\vspace*{.2cm}

We would like to thank S. Parvizi for useful comments and discussions.

\end{document}